\documentclass[a4paper]{jpconf}

\usepackage[english]{babel} 

\usepackage[caption=false, font=footnotesize, justification=RaggedRight]{subfig}

\usepackage[final]{graphicx}

\usepackage{subfig}

\usepackage[
   centertags, 
   sumlimits,  
   intlimits,  
   namelimits, 
]{amsmath} %
\usepackage{amssymb}

\usepackage[%
	pdftitle={Investigation of jet quenching and elliptic ﬂow within a pQCD-based partonic transport model},%
	pdfauthor={Oliver Fochler, Zhe Xu, and Carsten Greiner},%
	pdfsubject={Investigation of jet quenching and elliptic ﬂow within a pQCD-based partonic transport model},
	pdfstartview=FitH,
	pdfpagemode=UseNone,
	bookmarksopen=true
	]{hyperref}

\usepackage[all]{hypcap} 

\usepackage{dcolumn} 

\usepackage{multirow} 

\usepackage{microtype} 

\makeatletter
\def\tagform@#1{\maketag@@@{\ignorespaces#1\unskip\@@italiccorr}}
\let\orgtheequation\theequation
\def\theequation{(\orgtheequation)}
\makeatother

\newcommand{\beq}{\begin{equation}}
\newcommand{\eeq}{\end{equation}}

\graphicspath{{./}}
\DeclareGraphicsExtensions{.eps}

\usepackage{fancyhdr}
\fancypagestyle{plain}{%
\fancyhf{} 
\fancyhead[R]{\thepage} 

}
\begin{document}
\title{Investigation of jet quenching and elliptic flow within a pQCD-based partonic transport model}

\author{Oliver Fochler$^1$, Zhe Xu$^{2,1}$, and Carsten Greiner$^1$}

\address{$^1$ Institut f\"ur Theoretische Physik, Johann Wolfgang 
Goethe-Universit\"at Frankfurt, Max-von-Laue-Str. 1, 
D-60438 Frankfurt am Main, Germany}
\address{$^2$ Frankfurt Institute for Advanced Studies, Ruth-Moufang-Str. 1, D-60438 Frankfurt am Main, Germany}

\ead{fochler@th.physik.uni-frankfurt.de}

\begin{abstract}
The partonic transport model BAMPS (a Boltzmann approach to multiparton scatterings) is employed to investigate different aspects of heavy ion collisions within a common framework based on perturbative QCD. This report focuses on the joint investigation of the collective behavior of the created medium and the energy loss of high--$p_{T}$ gluons traversing this medium. To this end the elliptic flow and the nuclear modification factor of gluons in heavy ion collisions at 200 AGeV are simulated with BAMPS.

The mechanism for the energy loss of high energy gluons within BAMPS is studied in detail. For this, purely elastic interactions are compared to radiative processes, $gg \rightarrow ggg$, that are implemented based on the matrix element by Gunion and Bertsch. The latter are found to be the dominant source of energy loss within the framework employed in this work. 
\end{abstract}

\section{Introduction}
Experiments at the Relativistic Heavy Ion Collider (RHIC) have established that jets with high transverse momenta are suppressed in $Au + Au$ collisions with respect to a scaled $p + p$ reference \cite{Adler:2002xw,Adcox:2001jp}. This quenching of jets is commonly attributed to energy loss on the partonic level as the hard partons produced in initial interactions are bound to traverse the hot medium, the quark--gluon plasma (QGP), created in the early stages of the heavy ion collision. Another major discovery has been the strong collective flow of the created medium \cite{Adler:2003kt, Adams:2003am}, that is usually quantified in terms of the Fourier parameter $v_{2}$ and in this context often referred to as elliptic flow.

Due to the large momentum scales involved the energy loss of partonic jets can be treated in terms of perturbative QCD (pQCD) and most theoretical schemes attribute the main contribution to partonic energy loss to radiative processes \cite{Zakharov:1996fv, Baier:1996sk,Baier:1998yf,Gyulassy:2000er,Jeon:2003gi,Salgado:2003gb,Wicks:2005gt}. The bulk properties of the medium, such as the elliptic flow, are on the other hand usually investigated within hydrodynamical models. The comparison of hydrodynamic calculations to data indicates that the viscosity of the QGP is quite small \cite{Romatschke:2007mq}, possibly close to the conjectured lower bound $\frac{\eta}{s} = \frac{1}{4 \pi}$ from a correspondence between conformal field theory and string theory in an Anti-de-Sitter space \cite{Kovtun:2004de}.

It is a major challenge to combine these two aspects, jet physics on the one and bulk evolution on the other hand, within a common framework. Recently the efforts to combine pQCD-based jet physics with hydrodynamic modeling of the medium have been intensified, for instance results from hydrodynamical simulations are used as an input for the medium evolution in jet--quenching calculations (see \cite{Bass:2008rv} for an overview) and as ingredients in Monte Carlo event generators \cite{Schenke:2009gb}. However, these approaches still treat medium physics and jet physics in the QGP on very different grounds. Moreover, so far no schemes are available that cover the full dynamics of the interplay between jets and the medium.

Partonic transport models might provide means to investigate bulk properties of the QGP and high--energy parton jets within a common physical framework automatically including the full dynamics of the system evolution. In previous publications \cite{Fochler:2008ts, Fochler:2010wn} we have explored the capabilities of the transport model BAMPS (a Boltzmann approach to multiparton scatterings) with this goal in mind.

\section{The transport model BAMPS} \label{sec:BAMPS}

BAMPS \cite{Xu:2004mz,Xu:2007aa} is a microscopic transport model aimed at simulating the QGP stage of heavy ion collisions via pQCD interactions consistently including parton creation and annihilation processes. Partons within BAMPS are treated as semi-classical and massless Boltzmann particles and at this stage the model is limited to gluonic degrees of freedom. Thus $N_{f}=0$ is understood throughout this work. The strong coupling is fixed to $\alpha_{s}=0.3$. Please also note that the term ``jet'' as used throughout this paper refers to a single gluon with high energy that traverses the medium and does thus not fully coincide with the experimental notion.

The interactions between partons are based on leading order pQCD matrix elements from which transition probabilities are computed. These are used to sample the interactions of particles in a stochastic manner \cite{Xu:2004mz}. The test particle method is introduced to reduce statistical fluctuations.
For elastic interactions of gluons, $gg \leftrightarrow gg$, we use the Debye screened cross section in small angle approximation $\frac{d\sigma_{gg\to gg}}{dq_{\perp}^2} = \frac{9\pi\alpha_{s}^{2}}{(\mathbf{q}_{\perp}^2+m_D^2)^2}$. The Debye screening mass is computed from the local particle distribution $f=f(p,x,t)$ via $ m_{D}^{2} = d_G \pi \alpha_s \int \frac{d^3p}{(2\pi)^3} \frac{1}{p} N_c f$, where $d_G = 16$ is the gluon degeneracy factor for $N_c = 3$.

Inelastic $gg \leftrightarrow ggg$ processes are treated via the Gunion-Bertsch matrix element \cite{Gunion:1981qs}
\begin{equation} \label{eq:gg_to_ggg}
\left|\mathcal{M}_{gg \to ggg}\right|^2 = \frac{72 \pi^2 \alpha_s^2 s^2}{(\mathbf{q}_{\perp}^2+m_D^2)^2}\,
 \frac{48 \pi \alpha_s \mathbf{q}_{\perp}^2}{\mathbf{k}_{\perp}^2 [(\mathbf{k}_{\perp}-\mathbf{q}_{\perp})^2+m_D^2]}
\Theta\left( \Lambda_g - \tau \right)
\text{,}
\end{equation}
where $\mathbf{q}_{\perp}$ and $\mathbf{k}_{\perp}$ denote the perpendicular components of the momentum transfer and of the radiated gluon momentum in the center of momentum (CM) frame of the colliding particles, respectively. Detailed balance between gluon multiplication and annihilation processes is ensured by the relation $\left| \mathcal{M}_{gg \rightarrow ggg} \right|^{2} = d_{G} \left| \mathcal{M}_{ggg \rightarrow gg} \right|^{2}$.

The Theta function in \ref{eq:gg_to_ggg} is an effective implementation of the LPM (Landau, Pomeranchuk, Migdal) effect \cite{Migdal:1956tc}, that describes coherence effects in multiple bremsstrahlung processes. This interference effect cannot be incorporated directly into a semi--classical microscopic transport model such as BAMPS, so the cut--off $\Theta\left( \Lambda_g - \tau \right)$ ensures that successive $gg \rightarrow ggg$ processes are independent of each other. $\tau$ is the formation time of the gluon emitted with transverse momentum $k_{\perp}$ and $\Lambda_{g}$ denotes the mean free path, i.e. the time between successive interactions, of the parent gluon. When comparing the formation time to the mean free path of the parent gluon special attention needs to be paid to the frames of reference. This ultimately renders the cut--off dependent on the boost $\vec{\beta}$ between the plasma rest frame and the center of momentum (CM) frame in which \ref{eq:gg_to_ggg} is evaluated \cite{Fochler:2010wn} and numerically further complicates the calculations.

\section{Gluon jets in a static medium}

In order to obtain baseline calculations and to gain a better understanding of the energy loss mechanisms in BAMPS, we study the evolution of high energy gluons in a simplified brick setup, i.e. the propagation within a static thermal medium with fixed temperature $T$. Also, for reasons of computing time, possible effects of the propagating jet on the medium are neglected in most computations presented in this section. For example, the mean energy loss per unit path length $dE/dx$ is then calculated as follows ($c=1$)
\begin{equation}
\frac{dE}{dx} = \frac{dE}{d(ct)} = \sum_{i} \langle \Delta E^{i} \rangle R^{i} 
\end{equation}
where $i$ denotes the interaction type ($gg \rightarrow gg$, $gg \rightarrow ggg$ and $ggg \rightarrow gg$) and $R^{i}$ is the interaction rate for process $i$ and $\langle \Delta E^{i} \rangle$ is the mean energy loss in a single collision of type $i$.

\begin{figure}[tbh]
  \centering
  \begin{minipage}[t]{0.45\textwidth}
    \includegraphics[width=\linewidth]{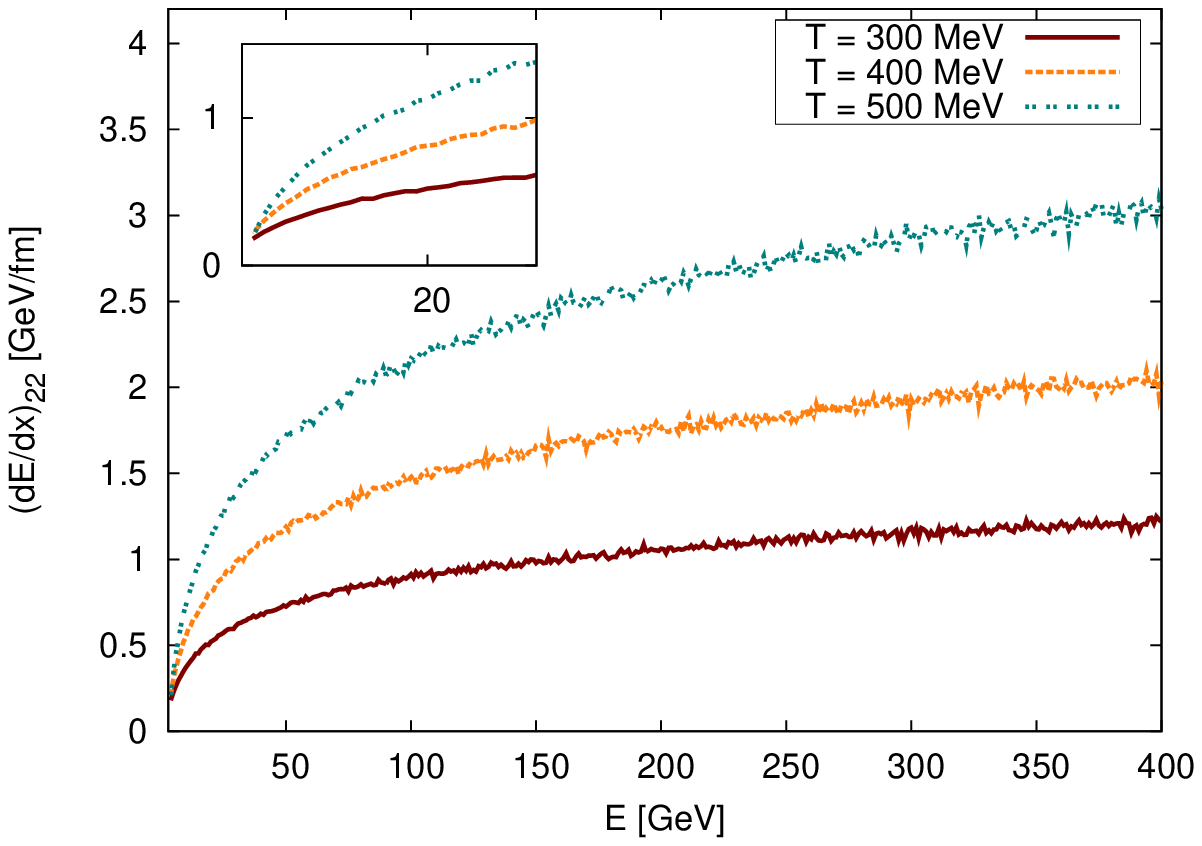}
  \end{minipage}
  \begin{minipage}[t]{0.45\textwidth}
    \includegraphics[width=\linewidth]{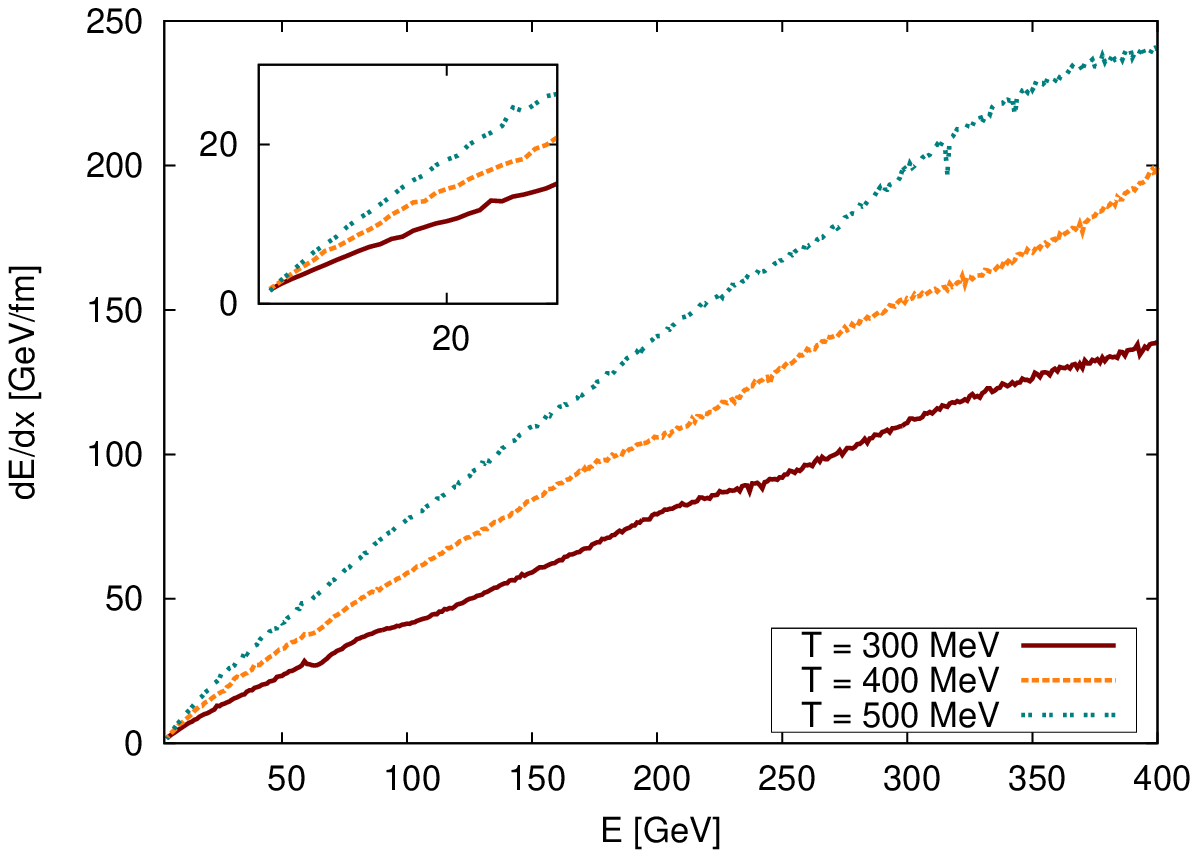}
  \end{minipage}
  \label{fig:dEdx}
  \caption{Differential energy loss of a gluon jet in a static and thermal medium of gluons at $T=400\,\mathrm{MeV}$, $T=500\,\mathrm{MeV}$ and $T=600\,\mathrm{MeV}$.\newline
  Left panel: Elastic interactions only. Right panel: Including $gg \leftrightarrow ggg$ processes.}
\end{figure}

The differential energy loss from elastic interactions as shown in the left panel of Fig. \ref{fig:dEdx} exhibits the expected (see \cite{Wicks:2005gt} for an overview) logarithmic dependence on the jet energy $E$ and the dominant quadratic dependence on the medium temperature $T$, $\left.\frac{dE}{dx}\right|_{2\rightarrow 2} \propto C_{R} \pi \alpha_{s}^{2} T^{2} \ln \left( \frac{4 E T}{m_{D}^{2}} \right)$, where $C_{R}$ is the quadratic Casimir of the propagating jet, $C_{R} = C_{A} = N_{c}$ for gluons. For $T = 400\,\mathrm{MeV}$ and $E = 50\,\mathrm{GeV}$ we find an elastic energy loss of  $\left.\frac{dE}{dx}\right|_{2\rightarrow 2} \approx 1.2\,\mathrm{GeV}/\mathrm{fm}$ that increases to $\left.\frac{dE}{dx}\right|_{2\rightarrow 2} \approx 2\,\mathrm{GeV}/\mathrm{fm}$ at $E=400\,\mathrm{GeV}$.

The energy loss caused by radiative $gg \rightarrow ggg$ interactions in BAMPS is much stronger and by far dominates the total differential energy loss (right panel of Fig. \ref{fig:dEdx}) of a high energy gluon. The differential energy loss from $gg \rightarrow ggg$ is rising almost linearly with the energy, for example resulting in a total $dE/dx \approx 32.6\,\mathrm{GeV}/{\mathrm{fm}}$ at $E = 50\,\mathrm{GeV}$ and $T=400\,\mathrm{MeV}$.

\begin{figure}[tbh]
  \centering
  \label{fig:sigma_and_deltaE}
  \begin{minipage}[t]{0.45\textwidth}
    \includegraphics[width=\linewidth]{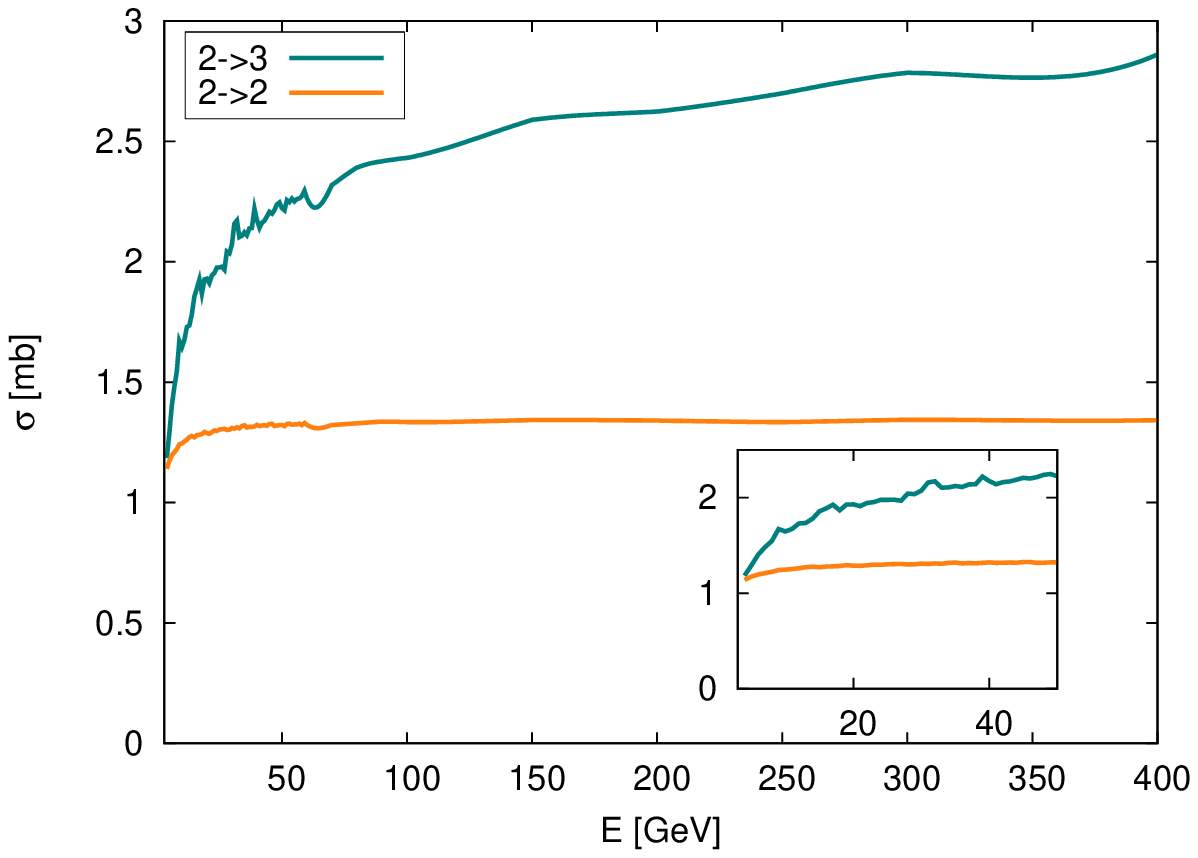}
  \end{minipage}
  \begin{minipage}[t]{0.45\textwidth}
    \includegraphics[width=\linewidth]{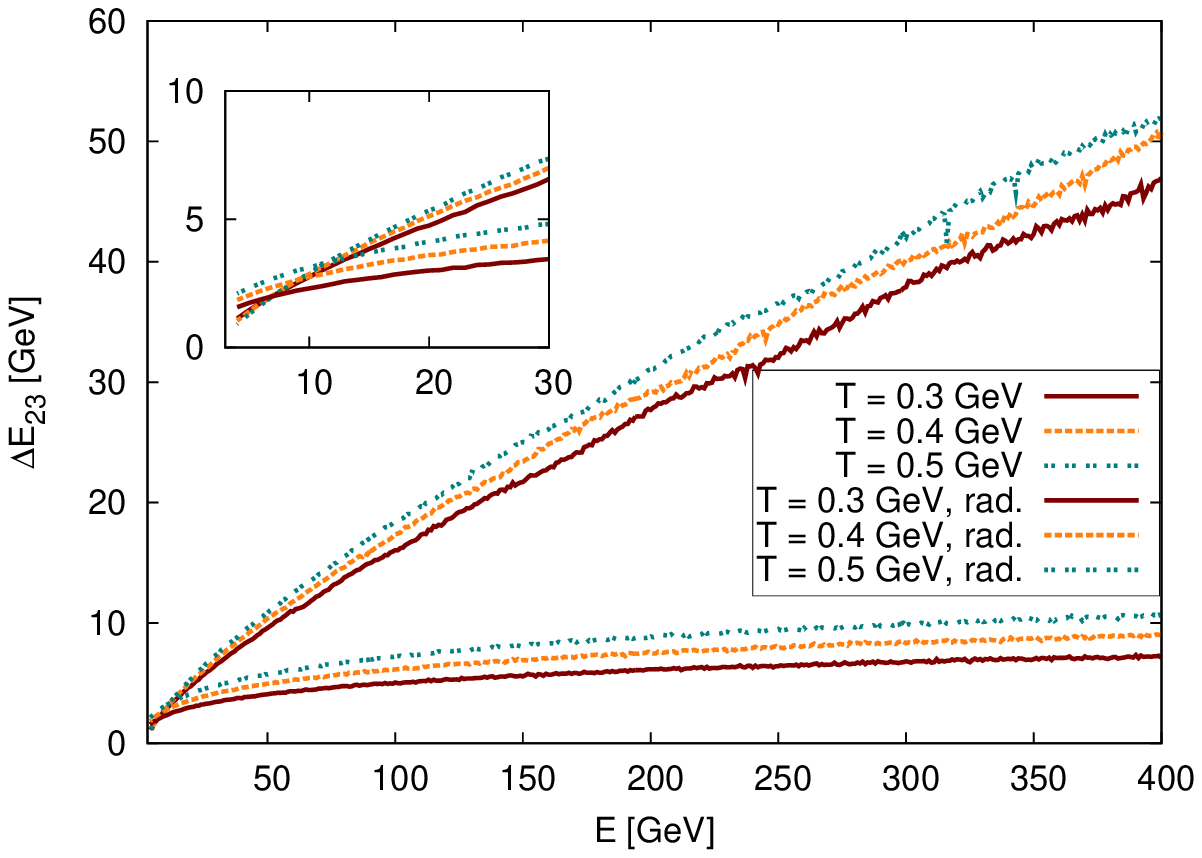}
  \end{minipage}
  \caption{Left panel: Cross sections for $gg \rightarrow gg$ ($2\rightarrow 2$) and $gg \rightarrow ggg$ ($2 \rightarrow 3$) processes.\newline
  Right panel: Mean energy loss $\langle \Delta E_{23} \rangle$ in a single $gg \rightarrow ggg$ process. Upper group of lines: $\Delta E = E^{\text{in}} - \max \left( E_{1}^{\text{out}}, E_{2}^{\text{out}}, E_{3}^{\text{out}} \right)$. Lower group of lines (labeled ``rad.''): $\Delta E = \omega$, with $\omega$ being the energy of the radiated gluon.\newline
  All quantities given as a function of the energy of the gluon jet inside a thermal medium with $T=400\,\mathrm{MeV}$.}
\end{figure}

The large differential energy loss in $gg \rightarrow ggg$ processes, however, is not governed by excessively strong cross sections. The individual cross sections increase only slowly with the jet energy as seen in the left panel of Fig. \ref{fig:sigma_and_deltaE}. For instance at $E=50\,\mathrm{GeV}$ and $T=400\,\mathrm{MeV}$ a gluon jet interacts with cross sections  $\left< \sigma_{gg \rightarrow gg} \right> \approx 1.3\,\mathrm{mb}$ and $\left< \sigma_{gg \rightarrow ggg} \right> \approx 3.5\,\mathrm{mb}$. This emphasizes that BAMPS does indeed operate with reasonable partonic cross sections based on pQCD matrix elements.

Rather it is the mean energy loss per single radiative process, $\langle \Delta E_{23} \rangle$, shown in the right panel of Fig. \ref{fig:sigma_and_deltaE}, that determines the magnitude of the differential energy loss $dE/dx$. It is noteworthy that the strong and linear rise in the energy loss due to $gg \rightarrow ggg$ is only present when identifying the outgoing gluon with the highest energy as the jet gluon and thus using $\Delta E = E^{\text{in}} - \max \left( E_{1}^{\text{out}}, E_{2}^{\text{out}}, E_{3}^{\text{out}} \right)$. This is the most natural choice and is employed throughout all calculations in this work. The average energy $\omega$ of the radiated gluon, however, is rising much slower with the jet energy. This is due to the fact that the energy is distributed among three outgoing particles, the gluon emitted with energy $\omega$ being only one of them. See \cite{Fochler:2010wn} for an in--depth discussion on how the complex interplay between kinematics, Gunion--Bertsch matrix element and phase space restriction imposed by the effective LPM cut--off gives rise to a fat tail in the $\Delta E_{23}$ distribution, that ultimately leads to large mean values $\langle \Delta E_{23} \rangle$.

\begin{figure}[tbh]
  \centering
  \begin{minipage}[t]{0.45\textwidth}
    \includegraphics[width=\linewidth]{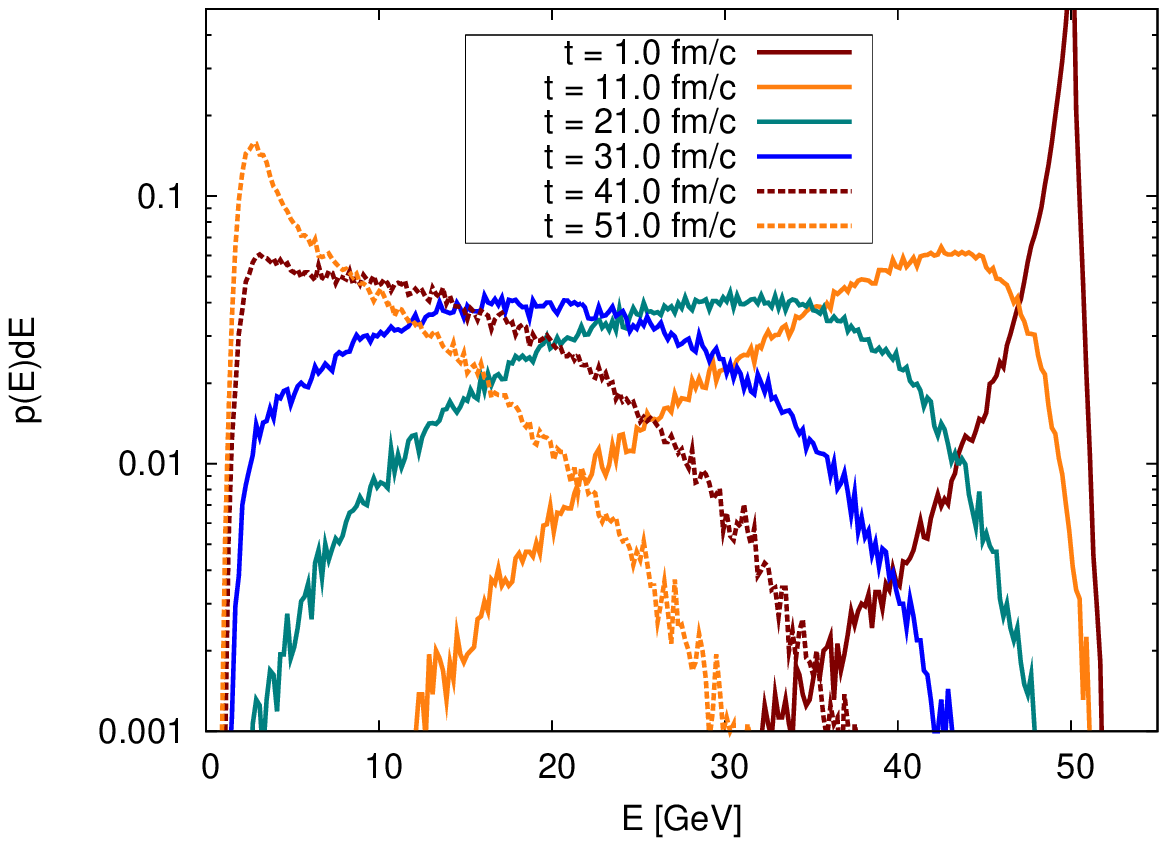}
  \end{minipage}
  \begin{minipage}[t]{0.45\textwidth}
    \includegraphics[width=\linewidth]{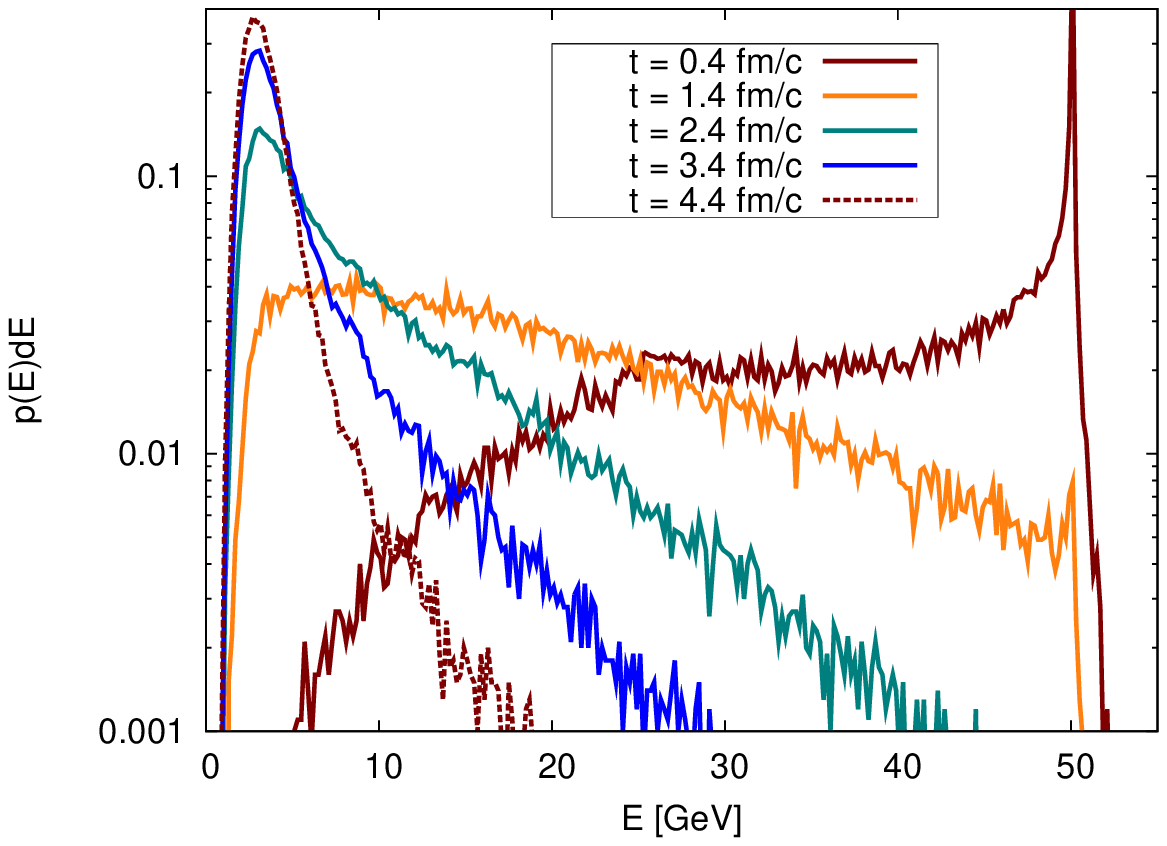}
  \end{minipage}
  \label{fig:energy_evolution}
  \caption{Time evolution of the energy distribution of a gluon jet that traverses a static and thermal medium of gluons ($T=400\,\mathrm{MeV}$). The initial ($t=0\,\mathrm{fm/c}$) energy of the gluon jet is $E_{0}=50\,\mathrm{GeV}$.\newline
  Left panel: Elastic interactions only. Right panel: Including $gg \leftrightarrow ggg$ processes.}
\end{figure}

The full time evolution of the energy distribution of a jet particle propagating through the medium obviously contains more detailed information than the mean energy loss per unit path length. Fig. \ref{fig:energy_evolution} shows  $\left. p(E)\,dE \right|_{t}$, the probability that a parton starting with $E(t=0\,\mathrm{fm/c})=E_{0} = 50\,\mathrm{GeV}$ has an energy $E \le E(t) < E + dE$ at a given time $t$.

For both $gg \rightarrow gg$ and $gg \rightarrow ggg$ the distribution of the jet energy induced by collisions with the constituents of the medium becomes rather broad. The distributions significantly differ from Gaussian shapes and a simple shift of the mean energy accompanied with momentum diffusion could not account for the behavior depicted in Fig. \ref{fig:energy_evolution}. A distinct peak at lower energies only re-emerges at very late times. The mean energy loss as depicted in Fig. \ref{fig:dEdx} is therefore a valuable observable but contains only limited information. It is noteworthy that there exists a finite probability for the jet to gain energy by collisions with the thermal gluons.

\begin{figure}[tbh]
  \centering
  \begin{minipage}[t]{0.45\textwidth}
    \includegraphics[width=\linewidth]{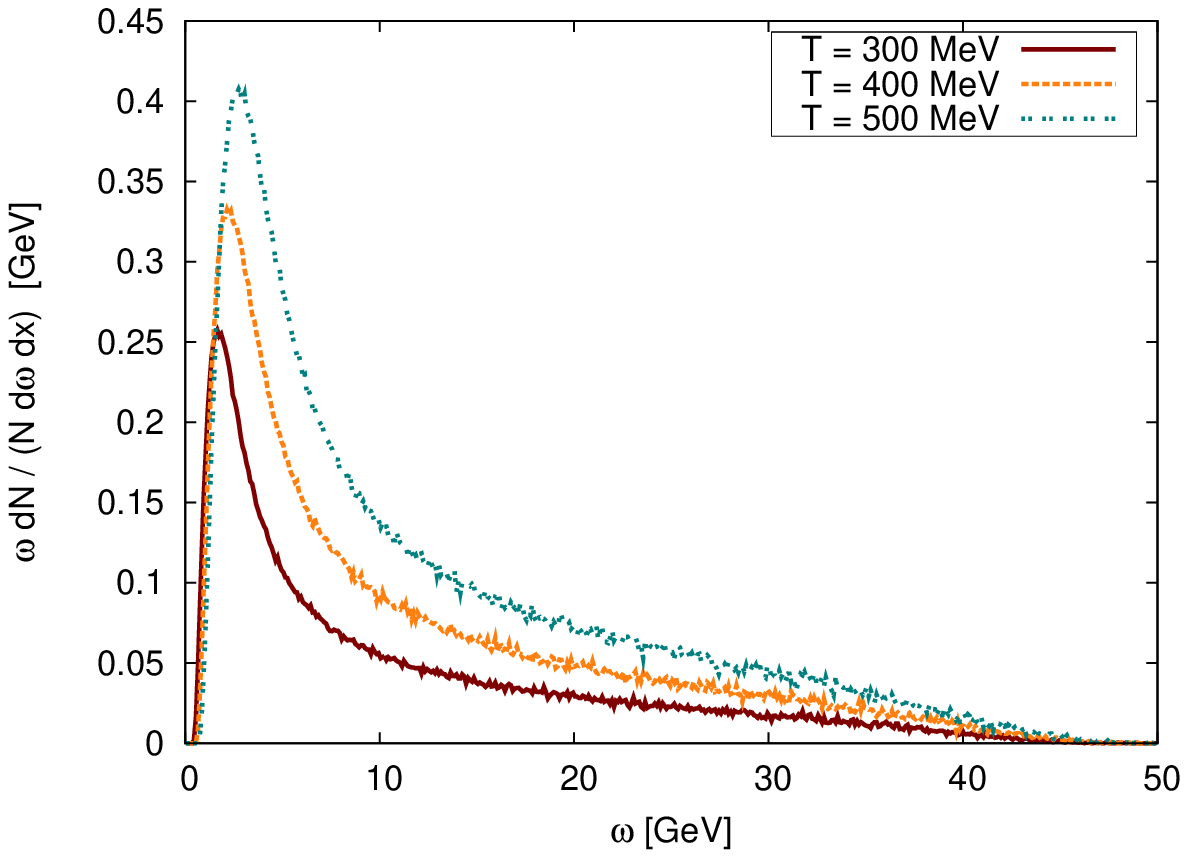}
  \end{minipage}
  \begin{minipage}[t]{0.45\textwidth}
    \includegraphics[width=\linewidth]{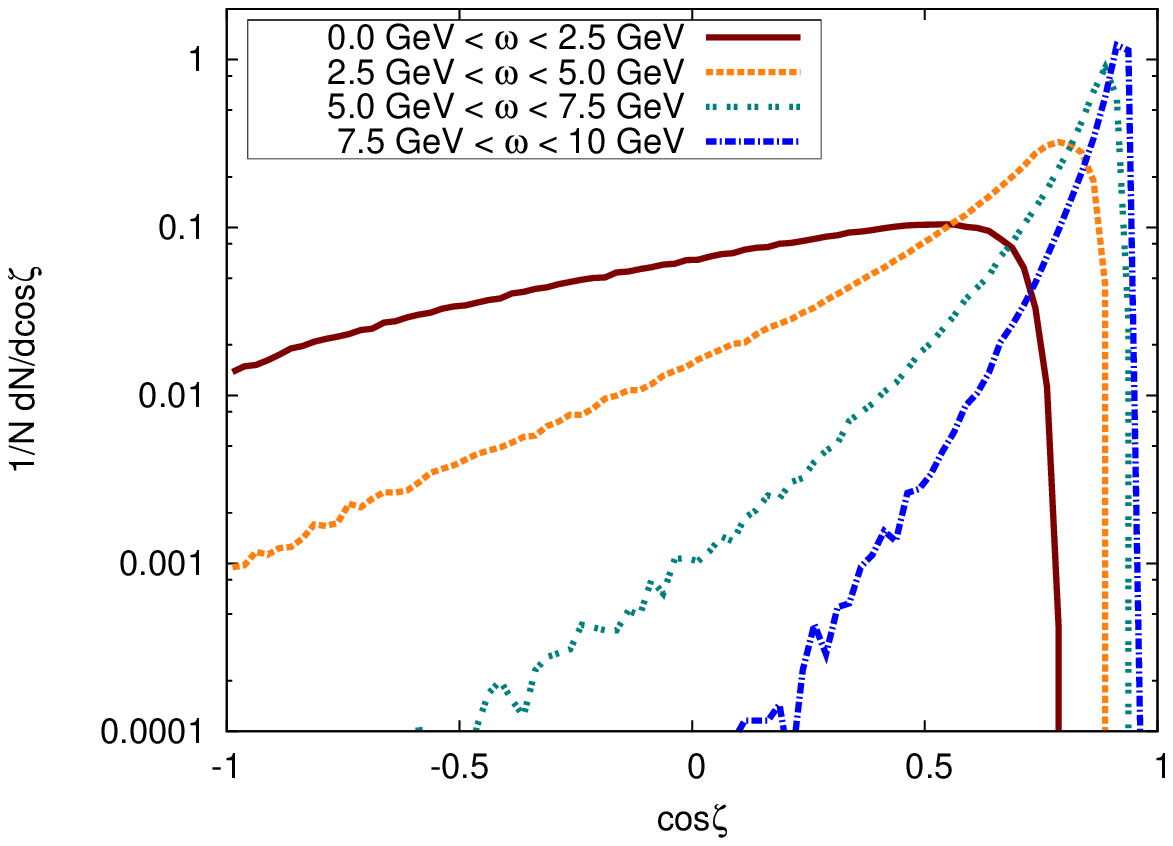}
  \end{minipage}
  \label{fig:rad_spectrum}
  \caption{Left panel: Energy spectrum $\omega \frac{dN}{N\,d\omega\,dx}$ of radiated gluons per energy interval $d\omega$ and distance $dx$. $\omega$ is the energy (lab frame) associated with the radiated gluon according to the Gunion--Bertsch matrix element. The energy of the gluon jet is $E=50\,\mathrm{GeV}$.\newline Right panel: Angular distribution of the radiated gluon in the lab frame with respect to the original jet direction for different energies $\omega$ of the radiated gluon. Jet energy $E = 50\,\mathrm{GeV}$, medium $T=400\,\mathrm{MeV}$.}
\end{figure}

Accompanying the above discussions, Fig. \ref{fig:rad_spectrum} shows the energy spectrum of gluons radiated in $gg \rightarrow ggg$ processes and the angular distribution of the radiated gluons for different ranges of their energy $\omega$. It is clearly visible that, due to the cut--off in transverse momentum \ref{eq:gg_to_ggg}, the gluons cannot be emitted at very forward angles, an effect that is more pronounced for low $\omega$. The spectra are peaked at energies $\omega \ll E$, with a small tail reaching out to high energies. With increasing temperature the peak of the spectrum shifts towards higher energies in an apparently linear way, favoring the emission of gluons with higher energies.

\section{Au+Au collisions at 200 AGeV}

\begin{figure}[tbh]
  \centering
  \begin{minipage}[t]{0.45\textwidth}
    \includegraphics[width=\linewidth]{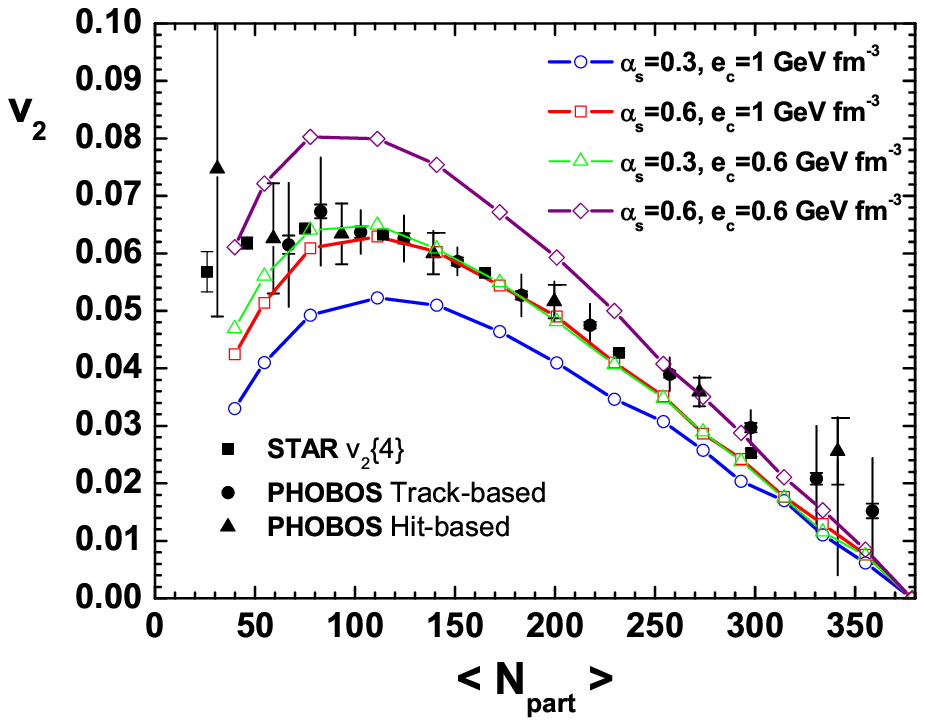}
  \end{minipage}
  \begin{minipage}[t]{0.45\textwidth}
    \includegraphics[width=\linewidth]{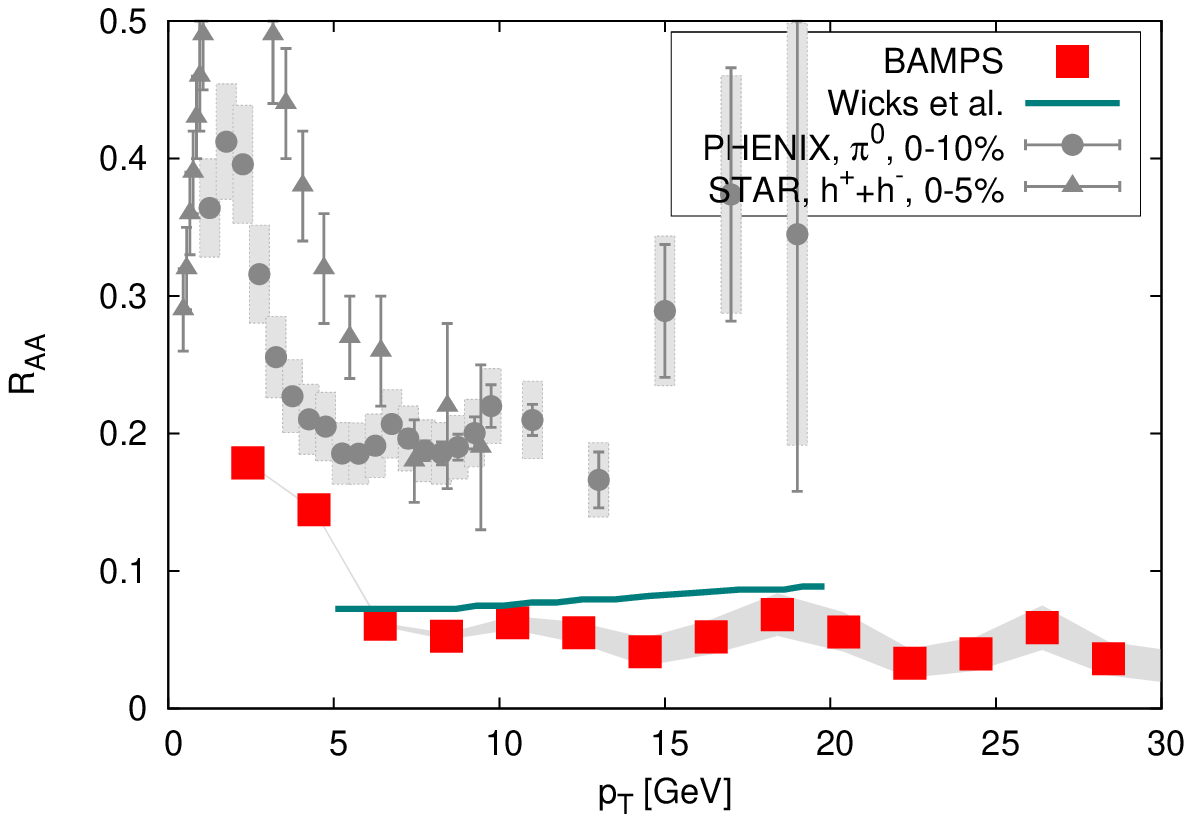}
  \end{minipage}
  \label{fig:v2_summary_RAA_central}
  \caption{Left panel: Elliptic flow $v_2$ as a function of the number of participants for Au+Au at 200~AGeV for different combinations of the strong coupling $\alpha_s$ and the critical energy density $\varepsilon_c$. See \cite{Xu:2008av} for more information. \newline Right panel: Gluonic $R_{AA}$ at midrapidity ($y \,\epsilon\, [-0.5,0.5]$) as extracted from simulations for central Au+Au collisions at 200~AGeV. For comparison the result from Wicks et al. \cite{Wicks:2005gt} for the gluonic contribution to $R_{AA}$ and experimental results from PHENIX \cite{Adare:2008qa} for $\pi^{0}$ and STAR \cite{Adams:2003kv} for charged hadrons are shown.}
\end{figure}

BAMPS has been applied to simulate elliptic flow and jet quenching at RHIC energies \cite{Fochler:2008ts}, for the first time using a consistent and fully pQCD--based microscopic transport model to approach both key observables on the partonic level within a common setup. The left panel of Fig. \ref{fig:v2_summary_RAA_central} shows that the medium simulated in the parton cascade BAMPS exhibits a sizable degree of elliptic flow in agreement with experimental findings at RHIC as established in \cite{Xu:2007jv, Xu:2008av}. And $\eta / s$ of the gluon matter in BAMPS has been shown to be small \cite{Xu:2007ns}.

For simulations of heavy ion collisions the initial gluon distributions are sampled according to a mini--jet model with a lower momentum cut-off $p_{0} = 1.4\,\mathrm{GeV}$ and a $K$--factor of $2$. The test particle method \cite{Xu:2004mz} is employed to ensure sufficient statistics and to allow for the resolution of adequate spatial length scales. The underlying nucleon-nucleon collisions follow a Glauber-model with a Wood-Saxon density profile and the results by Gl\"uck, Reya and Vogt \cite{Gluck:1994uf} are used as parton distribution functions. Quarks are discarded after sampling the initial parton distribution since currently a purely gluonic medium is considered. To model the freeze out of the simulated fireball, free streaming is applied to regions where the local energy density has dropped below a critical energy density $\varepsilon_{c}$ ($\varepsilon_{c} = 1.0\, \mathrm{GeV}/\mathrm{fm}^3$ unless otherwise noted). This setup has been successfully checked against experimental findings such as the distribution of transverse energy in rapidity and the flow parameter $v_{2}$ at various centralities in \cite{Xu:2007aa, Xu:2007jv, Xu:2008av}.  

The right panel of Fig. \ref{fig:v2_summary_RAA_central} shows the gluonic $R_{AA}$ simulated in BAMPS for central,  $b=0\,\mathrm{fm}$, collisions. It is roughly constant at $R_{AA}^{\mathrm{gluons}} \approx 0.053$ and in reasonable agreement with recent analytic results for the gluonic contribution to the nuclear modification factor $R_{AA}$ \cite{Wicks:2005gt}, though the suppression of gluon jets in BAMPS appears to be slightly stronger. We expect improved agreement in future studies when employing a carefully averaged $\langle b \rangle$ that will be better suited for comparison to experimental data than the strict $b=0\,\mathrm{fm}$ case.

\begin{figure}[tbh]
  \centering
  \begin{minipage}[t]{0.45\textwidth}
    \includegraphics[width=\linewidth]{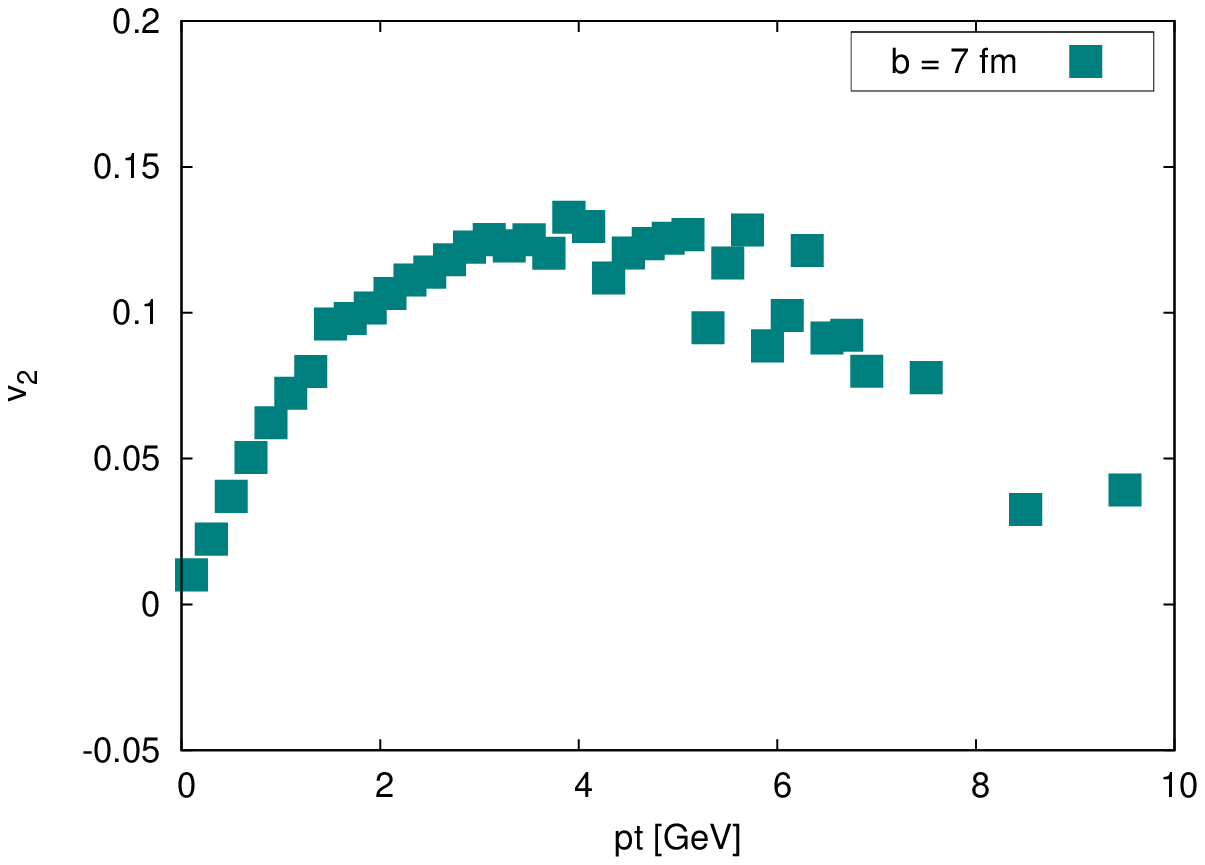}
  \end{minipage}
  \begin{minipage}[t]{0.45\textwidth}
    \includegraphics[width=\linewidth]{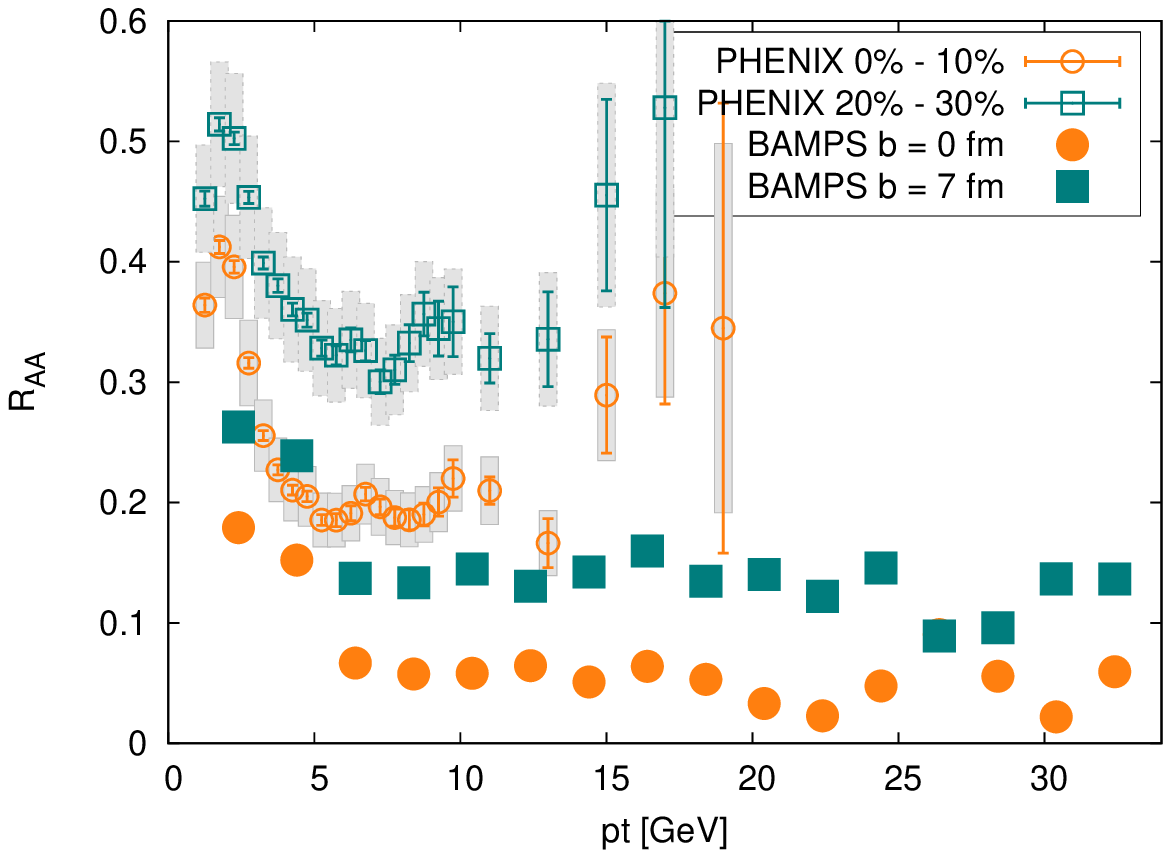}
  \end{minipage}
  \label{fig:v2_RAA_b7}
  \caption{Left panel: Elliptic flow $v_{2}$ for gluons in simulated Au+Au collisions at 200 AGeV with $b=7\,\mathrm{fm}$. $\varepsilon_c = 0.6\, \mathrm{GeV}/\mathrm{fm}^3$. \newline
  Right panel: Gluonic $R_{AA}$ as extracted from BAMPS simulations for $b=0\,\mathrm{fm}$ and $b=7\,\mathrm{fm}$, $\varepsilon_c = 1.0\, \mathrm{GeV}/\mathrm{fm}^3$. For comparison experimental results from PHENIX \cite{Adare:2008qa} for $\pi^{0}$ are shown for central ($0 \%$ - $10 \%$) and off--central ($20 \%$ - $30 \%$) collisions.}
\end{figure}

As a first step towards making more extensive comparisons with experimental data and analytic models possible, we have computed the gluonic $R_{AA}$ for non--central Au + Au collisions at the RHIC energy of $\sqrt{s} = 200 \mathrm{AGeV}$ with a fixed impact parameter $b=7\,\mathrm{fm}$, which roughly corresponds to $20 \%$ to $30 \%$ experimental centrality. The results is shown in the right panel of Fig. \ref{fig:v2_RAA_b7}.

A comparison in terms of the magnitude of the jet suppression for $b=7\,\mathrm{fm}$ is difficult since there are no published results for the gluonic contribution to $R_{AA}$ from analytic models available. Taking the ratio of the $b=7\,\mathrm{fm}$ to the $b=0\,\mathrm{fm}$ results as a rough guess indicates that the decrease in quenching is more pronounced in BAMPS compared to experimental data. The ratio of the nuclear modification factor between central ($0 \%$ - $10 \%$) and more peripheral ($20 \%$ - $30 \%$) collisions is $\left. R_{AA}\right|_{0 \% - 10 \%} / \left. R_{AA}\right|_{20 \% - 30 \%} \approx 0.6$ for the experimental data, while for the BAMPS results $\left. R_{AA}\right|_{b=0\,\mathrm{fm}} / \left. R_{AA}\right|_{b=7\,\mathrm{fm}} \approx 0.4$. However, the issue of detailed quantitative comparison needs to be re-investigated once light quarks and a fragmentation scheme are included into the simulations.

To complement the investigations of $R_{AA}$ at a non--zero impact parameter $b=7\,\mathrm{GeV}$, we have computed the elliptic flow parameter $v_{2}$ for gluons at the same impact parameter and extended the range in transverse momentum up to roughly $p_{T} \approx 10\,\mathrm{GeV}$, see left panel of Fig. \ref{fig:v2_RAA_b7}. For this calculation we have used a critical energy density $\varepsilon_{c} = 0.6\, \mathrm{GeV}/\mathrm{fm}^3$ in order to be comparable to previous calculations.

The $v_{2}$ of high--$p_{T}$ gluons is rising up to $p_{T} \approx 4\,\mathrm{GeV}$. Afterwards, from about $p_{T} \approx 5\,\mathrm{GeV}$ on, the elliptic flow slightly decreases again with $p_{T}$. This behavior is in good qualitative agreement with recent RHIC data \cite{Abelev:2008ed} that for charged hadrons shows $v_{2}$ to be rising up to $v_{2} \approx 0.15$ at $p_{T} \approx 3\,\mathrm{GeV}$ followed by a slight decrease.

\section{Summary}

We have computed the gluonic contribution to the nuclear modification factor $R_{AA}$ and the elliptic flow $v_{2}$ employing the pQCD based transport model BAMPS. This model provides means to investigate various characteristics of the evolution of the partonic medium created in heavy ion collisions, ranging from bulk properties to high--$p_{T}$ physics, consistently including the full dynamics of the system.

The gluonic $R_{AA}$ is found to be flat over a wide range in $p_{T}$ at $R_{AA} \approx 0.13$ in off--central events ($b=7\,\mathrm{GeV}$) and at $R_{AA} \approx 0.053$ in central events ($b=0\,\mathrm{GeV}$) for a critical energy density of $\varepsilon_{c} = 1.0\, \mathrm{GeV}/\mathrm{fm}^3$. Since BAMPS allows for the simultaneous investigation of high--$p_{T}$ observables and bulk properties of the medium, we have also studied the elliptic flow parameter for gluons up to roughly $10\,\mathrm{GeV}$ for Au+Au at $b=7\,\mathrm{fm}$. $v_{2}$ peaks at a $p_{T}\approx 4 \div 5\,\mathrm{GeV}$ and slowly drops towards larger transverse momenta.

In order to systematically investigate the energy loss of gluons as implemented in BAMPS we have studied the evolution of high energy gluons within thermal and static media of gluons. Inelastic $gg \rightarrow ggg$ processes are found to be the dominant source of energy loss for high energy gluons in computations within the BAMPS model resulting in a strong differential energy loss that rises almost linearly with the jet energy. The strong mean energy loss in $gg \rightarrow ggg$ processes is due to a heavy tail in the $\Delta E$ distribution for single interactions, caused by the phase space configurations of outgoing particles dictated by the Gunion-Bertsch matrix element in combination with the effective LPM cutoff \cite{Fochler:2010wn}.

The characteristics of the strongly interacting, but still fully pQCD based, medium within the BAMPS description will be studied in further detail in upcoming works including light quark degrees of freedom. While a consistent modeling of low--$p_{T}$ hadronization needs careful consideration, the application of fragmentation functions to the high--$p_{T}$ sector will be straightforward once light quarks are included and will allow for more direct comparison to hadronic observables. Also the application of BAMPS to heavy quark elliptic flow and quenching will provide further valuable insight and is underway \cite{Uphoff:2010sh}. Additionally the medium response to high--$p_{T}$ particles will be studied in more detail. It has already been demonstrated \cite{Bouras:2009nn} that BAMPS offers the ability to describe collective shock phenomena in a viscous hydrodynamic medium.

\section*{Acknowledgments}
This work has been supported by the Helmholtz International Center for FAIR within the framework of the LOEWE program launched by the State of Hesse. The simulations have been performed at the Center for Scientific Computing (CSC) at the Goethe University Frankfurt.

\section*{References}
\bibliography{fochler}

\end{document}